\documentclass[aps,prb,superscriptaddress,twocolumn,showpacs,floatfix]{revtex4-1}
\usepackage[dvipdfmx]{graphicx}
\usepackage{bmpsize}

\usepackage{bm}

\DeclareSymbolFont{newfont}{OML}{cmm}{m}{it} % Computer Modern math font
\DeclareMathSymbol{\Epsilon}{3}{newfont}{15} % Symbol 15

\usepackage{amssymb}

\usepackage{color}
\definecolor{darkgreen}{rgb}{0.0, 0.3, 0.0}
\definecolor{armygreen}{rgb}{0.29, 0.33, 0.13}

\newcommand{\nano}{\mathrm{n}}
\newcommand{\milli}{\mathrm{m}}
\newcommand{\meter}{\mathrm{m}}
\newcommand{\electronvolt}{\mathrm{eV}}
\newcommand{\volt}{\mathrm{V}}
\newcommand{\kelvin}{\mathrm{K}}
\newcommand{\angstrom}{\AA}

\newcommand{\SI}[2]{#1{{\;}#2}}

\newcommand{\enquote}[1]{"{#1}"}

\begin{document}

\title{Nanoelectronic devices based on twisted graphene nanoribbons}


%\titlerunning{Nanoelectronic devices based on twisted graphene nanoribbons}

\author{Marta Saiz-Bret\'{\i}n}
%\email{marta.saiz.bretin@ucm.es}
\affiliation{GISC, Departamento de F\'{\i}sica de Materiales, Universidad Complutense, E-28040 Madrid, Spain}
\author{Andrey V.~Malyshev}
\email[Corresponding author: ]{a.malyshev@fis.ucm.es}
\affiliation{GISC, Departamento de F\'{\i}sica de Materiales, Universidad Complutense, E-28040 Madrid, Spain}
\affiliation{Ioffe Physical-Technical Institute, St-Petersburg, Russia}
\author{Francisco Dom\'{\i}nguez-Adame}
\affiliation{GISC, Departamento de F\'{\i}sica de Materiales, Universidad Complutense, E-28040 Madrid, Spain}


%\institute{GISC, Departamento de F\'{\i}sica de Materiales, Universidad Complutense, E-28040 Madrid, Spain
%\email{marta.saiz.bretin@ucm.es}
%}


%\maketitle

\begin{abstract}
We argue that twisted graphene nanoribbons subjected to a transverse electric field can operate as a variety of nanoelectronic devices, such as tunable tunnel diodes with current-voltage characteristics controlled by the transverse field. Using the density-functional tight-binding method to address the effects of mechanical strain induced by the twisting, we show that the electronic transport properties remain almost unaffected by the strain in relevant cases and propose a simplified tight-binding model which gives reliable results. The transverse electric field creates a periodic electrostatic potential along the nanoribbon, resulting in a formation of a superlattice-like energy band structure and giving rise to different remarkable electronic properties. We demonstrate that if the nanoribbon geometry and operating point are selected appropriately, the system can function as a field-effect transistor or a device with nonlinear current-voltage characteristic manifesting one or several regions of negative differential resistance. The latter opens possibilities for applications such as an active element of nanoscale amplifiers, generators, and new class of devices with multiple logic states.
%\keywords{graphene \and  nanoribbons \and twist \and negative differential resistance}
\end{abstract}

\maketitle

\section{Introduction} \label{sec:intro}

Chiral conformations are very common in nature and can be found at practically all length scales. Because of their peculiar properties, they offer underlying technological solutions in a variety of areas extending from the macroscopic to the nanoscopic worlds. Helical structures can either self-assemble naturally or be fabricated. Several growth and fabrication techniques have been successfully used to produce different chiral systems \cite{Motojima1990,Amelinckx1994,Zhang1994,Prinz2000,Kong2003,Zhang2003,Zhang2004,Yang2005,Gao2005,Gao2014}. There has also been a considerable effort to study fundamental properties and applications of nanohelices recently; examples range from more traditional semiconductor systems~\cite{Kibis2005b, Kibis2007, Kibis2008, Downing2016} to macromolecules, such as $\alpha$-helices and the DNA~\cite{Klotsa2005,Malyshev2007}.

Potential applications of chiral systems include energy storage~\cite{Gao2006}, sensing~\cite{Hwang2013}, THz generation~\cite{Kibis2007b,Portnoi2008,Portnoi2009,Rosenau2009,Batrakov2010}, stretchable electronics~\cite{Xu2011}, or spin selectivity~\cite{Goehler11,Xie11,Gutierrez13,Diaz18}, to name a few. Furthermore, when subjected to a transverse electric field, the helical motion of a charge carrier in a chiral system can result in the appearance of superlattice properties~\cite{Kibis2005a}, giving rise to a variety of phenomena and potential applications, such as electrical signal amplification and terahertz generation by systems with the negative differential resistance~(NDR) or electromagnetic wave generation by quantum cascade lasers~\cite{Kazarinov1971,Faist1994}.

Recently, a range of methods has been put forward to obtain twisted graphene nanoribbons~\cite{Elias10,Khlobystov11,Chamberlain12,Zhang2014}, which opens up a new possible route to further exploit the induced superlattice properties of graphene based systems. Elastic and thermal response of nanostructured graphene can be significantly altered as compared to those of the bulk material~\cite{Hu09,Balandin11,Yang13,Saiz-Bretin17}. Numerical calculations show that carbon nanotubes remain almost straight even at $T=\SI{700}{\kelvin}$ while the typical conformation of a free-standing graphene nanoribbon (GNR) is fully random at this temperature~\cite{Bets09}. 
At lower temperatures, quantum mechanical effects become important: the charge density in edge atoms' orbitals is redistributed resulting in edge reconstruction which can be interpreted as an effective strain of bonds at the edge and give rise to different non-planar configurations~\cite{Shenoy08,Wang12}.
%
%In wider GNRs, twists vanish while some warping of the edges can  remain~\cite{Bets09,Shenoy08,Wang12}. 
%
Mechanical deformations, and particularly twists, can also be induced and controlled. Theoretical studies suggest that chemistry at the edges~\cite{Gunlycke10,Nikiforov14} or tilt grain boundaries~\cite{Liu15} can be used to induce twisting. At the same time, experiments show that fabrication of helical GNRs is possible, for example, by cutting carbon nanotubes laterally~\cite{Elias10} or using them as reactors~\cite{Chamberlain12,Khlobystov11} or by hydrogen doping of graphene nanoribbons~\cite{Zhang2014}. The feasibility of the GNR conformation control is a remarkable feature and a very promising tool for nanoelectronic applications. It has been demonstrated that the helical conformation affects electronic~\cite{Tang12,Sadrzadeh11,Xu15,Atanasov15}, electromechanical~\cite{Al-Aqtash13,Jia14}, mechanical~\cite{Cranford11}, magnetic~\cite{Yue14}, thermal~\cite{Wei14,Antidormi17} and thermoelectric~\cite{Liu14} properties. However, possibilities of control of physical properties of twisted GNRs have not been studies so extensively.
%In this regard, it was found that the energy gap in helical armchair GNRs~(AGNRs) can be modulated by the twist angles, whereas helical zig-zag GNRs~(ZGNRs) are metallic regardless of the twist~\cite{Sadrzadeh11,Xu15}. 

In this work, we first study the influence of deformations induced by twisting on the electronic properties of GNRs. To this end we use the well established density-functional based tight-binding (DFTB) method and demonstrate that the effects of deformations on the transport properties can be neglected in relevant cases. The latter justifies the usage of a much simpler tight-binding method throughout the rest of the paper for modeling of the electronic characteristics of twisted GNRs.  
%Then, a transverse electric field is applied and the electron transmission through the GNRs is studied. The transmission coefficient allows us to calculate the current-voltage characteristics using the Landauer-B\"{u}ttiker formalism in the fully ballistic regime.\cite{Datta97} This approximation is reasonable in view of the near-ballistic electron transport observed in graphene field-effect transistors at room temperature.\cite{Dragoman16} The helical conformation of the carbon atoms in twisted GNRs induces an additional periodic potential when a transverse electric field is applied. We show that the electric current flowing through the GNRs can be efficiently tuned by varying the magnitude of the field. Furthermore, we show that if the working point is carefully selected, the device manifests negative differential resistance and display N--shaped current-voltage characteristic.
Further, we show that the current-voltage characteristics of the system can be controlled by the transverse electrostatic field. They can be engineered in such a way that the system can operate either as a field-effect transistor or as a device with highly nonlinear N-shape current-voltage characteristic with one or several NDR regions.

\section{System, methodology, and model} \label{sec:model}

\begin{figure}[ht]
   \begin{center}
          \includegraphics[width=\linewidth]{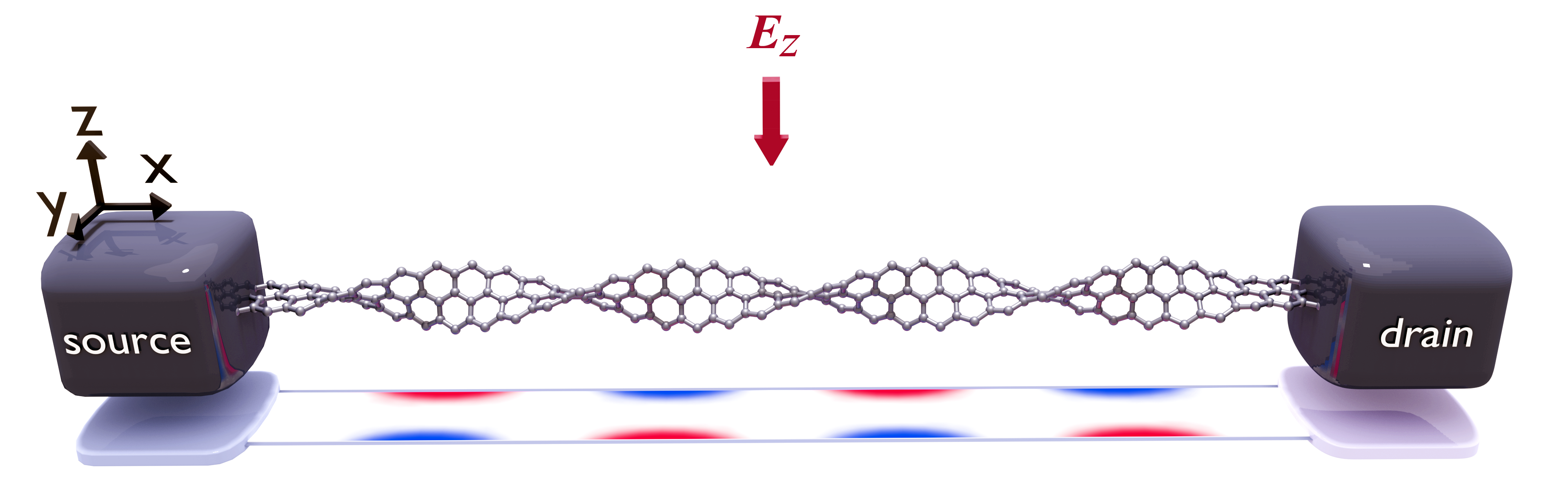}
   \end{center}
   \caption{Schematics of a $3$--ZGNR twisted four times, connected to source and drain leads, and subjected to the transverse electric field ${\bm E}_z$, applied along the $z$-axis. The map of the electrostatic potential landscape induced by the field is shown at the bottom of the plot, where red (blue) color represents higher (lower) electric potential.}
   \label{fig1}
\end{figure}

%\A{
The schematics of the considered system is shown in Figure~\ref{fig1}. The system comprises a GNR of length $L$ and width $W$, twisted $n$ times (each twist being by $180^{\circ}$ around the longitudinal symmetry $x$-axis), and connected to a pair of source and drain leads. The system is biased by the source-drain voltage $V_\mathrm{SD}$ and subjected to the homogeneous transverse electrostatic field ${\bm E}_z$ applied in the $z$-direction. The transverse field induces a periodic electrostatic potential in the twisted GNR as shown schematically at the bottom of the Figure~\ref{fig1}, where red (blue) color represents higher (lower) potential. The width of a GNR is commonly specified in terms of the number of dimer lines, $N$, in the transverse direction. Hereafter we use the notation $N$--AGNR and $N$--ZGNR for graphene nanoribbons with $N$ dimer lines and armchair or zig-zag edges, respectively. 
%}%\A

Our methodology is the following. First, we use the density-functional based tight-binding method, as implemented in the DFTB+ software package (see Ref.~\citenum{Aradi07} and references therein) with the parameter set mio-1-1~\cite{Elstner98}, to model the structural relaxation of twisted GNRs. 
%
% The DFTB method can be understood as an approximate density functional theory scheme which also uses ideas of the semiempirical tight-binding methods.\cite{Seifert07} Its advantage lies in the combination of accuracy and numerical efficiency. 
%
%\A{%
The DFTB method has been applied successfully for a large variety of problems in physics, chemistry, biology and material science~\cite{Frauenheim2002}, in particular, graphene structures~\cite{Zobelli11,Sevincli13,Liao17,Medrano-Sandonas18}, demonstrating good agreement with experimental data and results obtained with more accurate \emph{ab initio\/} methods. The method allows us to calculate positions of atoms and chemical bond lengths in the relaxed structure, which we use further in our calculations of the transmission spectrum of the system. To this end, we use the standard tight-binding Hamiltonian of a single electron in the $\pi$-orbitals of C atoms within the nearest-neighbor approximation
%}%\A
%
\begin{equation}
\mathcal{H}=\sum_{i}\varepsilon_i|i \rangle\langle i| 
-\sum_{\langle i,j\rangle}t_{ij}|i\rangle\langle j|\ ,
\label{ham}
\end{equation}
where $\varepsilon_i$ is the position-dependent energy of the orbital state $|i \rangle$ and $t_{ij}$ is the hopping energy. The second sum is restricted to nearest-neighbor atoms only. To account for effects of the bond strain, we use the conventional dependence of the hopping energy $t_{ij}$ on the bond length $d_{ij}$ (see Ref.~[\citenum{Ribeiro09}] and references therein)
\begin{equation}
t_{ij}=t_0\,\exp\left(-\beta \Epsilon_{ij}\right)\ ,
\qquad
\Epsilon_{ij}= (d_{ij}-a_0)/a_0\ ,
\label{hopping}
\end{equation}
where $t_0=\SI{2.7}{\electronvolt}$ is the hopping energy in unstrained graphene and $\beta$ is a dimensionless parameter in the range $3-4$~\cite{Ribeiro09} and $a_{0}=\SI{0.142}{\nano\meter}$ is the equilibrium bond length in graphene. . We use $\beta=4$ to account for the strongest possible dependence.

%\A{
In the presence of the source-drain bias $V_\mathrm{SD}$ and the transverse electric field ${\bm E}_z$, the orbital energies have the form 
\begin{equation}
\varepsilon_i=-e\,\bm E(\bm r_i)\cdot \bm r_i\ ,
\label{site-energy-real}
\end{equation}
where $-e$ is the electron charge, $\bm r_i$ is the position vector of the $i$-th atom in the relaxed structure, and $\bm E(\bm r_i)$ is the full electric field at the atom position. However, as we demonstrate in the next section, the effects related to the structural relaxation can be neglected in relevant cases and the following simple approximation of the orbital energy can be used
\begin{equation}
\varepsilon_i=-e\,V_\mathrm{SD}\left(\frac{x_i}{L}\right)-e\,{E}_z\,y_i\,\sin\left(\frac{\pi\,x_i}{\lambda}\right)\ .
\label{site-energy}
\end{equation}
Here $0\leq x_i\leq L$ and $-W/2\leq y_i \leq W/2$ are the coordinates
of the $i$-th C atom in the {\it pristine} GNR while $\lambda=L/n$ is the
twist length.  Strictly speaking, the transverse component of the full
electric field should be corrected for the polarization of the GNR, but
recent self-consistent calculations of the energy structure of GNRs
subjected to a transverse electric field show that the polarization effect
can be neglected up to the field intensities on the order of
$E_z=\SI{0.1:0.2}{\volt/\angstrom}$~\cite{Alaei13}.  Smaller magnitudes of the
electric field are used in our study and, therefore, the renormalization due
to the GNR polarization is neglected.  Finally, the leads are modeled in the
standard way: as semi-infinite planar GNRs (in the $x$-$y$ plane with $x<0$
and $x>L$) with zero orbital energy.

The phase coherence length of electronic states in graphene can be very
large even at room temperature~\cite{Dragoman16} and therefore we assume
that electron transport is ballistic and compute wave functions and
transmission coefficient using the quantum transmission boundary
method~\cite{Lent90,Ting92}, combined with the effective transfer matrix
method~\cite{Schelter10} (see Ref.~\citenum{Munarriz14} for further details
on the calculation method).  
%}%\A

\section{Structural relaxation effects}

%\A{
Twisted conformation of a GNR imposes a purely geometrical change of inter-atomic distances with respect to the pristine GNR case. The final atomic positions are determined by the helical geometry and structural relaxation occurring due to redistribution of the electronic density in atomic orbitals. In order to model these effects in a twisted GNR we used the DFTB method, in which the relaxation was performed by the conjugate gradient method until the absolute value of the inter-atomic forces were below $10^{-5}$ atomic units (the extremes of the ribbon were kept fixed in the simulation while the edges were H-passivated). Then, atom positions $\bm r_i$ and bond lengths $d_{ij}$ were obtained and further the strain of the ${ij}$--bond was calculated as $\Epsilon_{ij}=(d_{ij}-a_0)/a_0$.

\begin{figure}[hb]
   \begin{center}
       \includegraphics[width=0.9\linewidth]{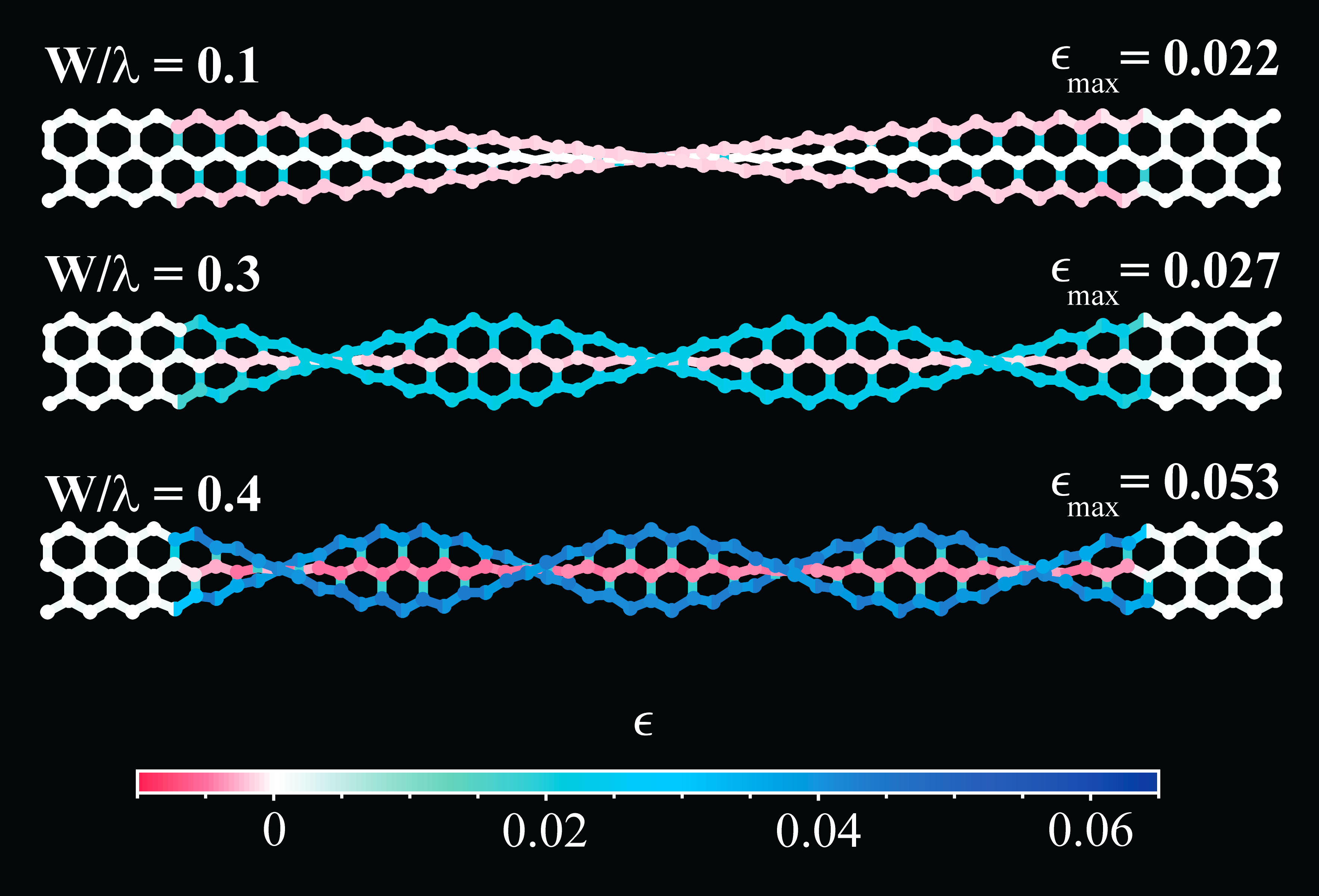}
   \end{center}
   \caption{Strain distribution in twisted and relaxed $\SI{4.4}{\nano\meter}$ long $3$--ZGNRs for different number of twists and values of the torsion coefficient $W/\lambda$ which is specified by each image together with the maximum value of the strain $\Epsilon_\mathrm{max}$. The passivating H atoms are not shown for clarity.}
   \label{fig2}
\end{figure}

Figure~\ref{fig2} shows examples of the strain distribution in relaxed structures of a $\SI{4.4}{\nano\meter}$ long $3$--ZGNR twisted $1$, $3$, and $4$ times. For convenience we introduce the dimensionless torsion coefficient, $\Upsilon=W/\lambda=n\,(W/L)$, which combines all the geometrical parameters defining a twisted ribbon and turns out to be a very useful characteristic of the system, as we argue below. For the lowest considered value of the torsion coefficient (upper image of Figure~\ref{fig2}) the strain at the edges is still slightly negative, indicating that the corresponding bonds are shorter. This result agrees qualitatively with previous {\emph{ab-initio}} calculations~\cite{Okada2008,Jun2008} where it was found that the edges are under effective compression due to the charge density redistribution. As the number of twists increases the edge bonds become stretched (see the two lower images in the figure). The latter can be understood as a purely \enquote{geometrical} effect: if the ribbon width is kept constant while it is twisted more and more times, the total edge length grows, resulting in the increase of each edge bond length. As the figure shows, in the latter case the maximum strain is located at the edges of a ribbon. As the torsion increases further, the strain becomes larger and eventually the edge bonds break (this case is not shown here). 
%}%\A

\begin{figure}[ht]
   \begin{center}
       \includegraphics[width=0.8\linewidth]{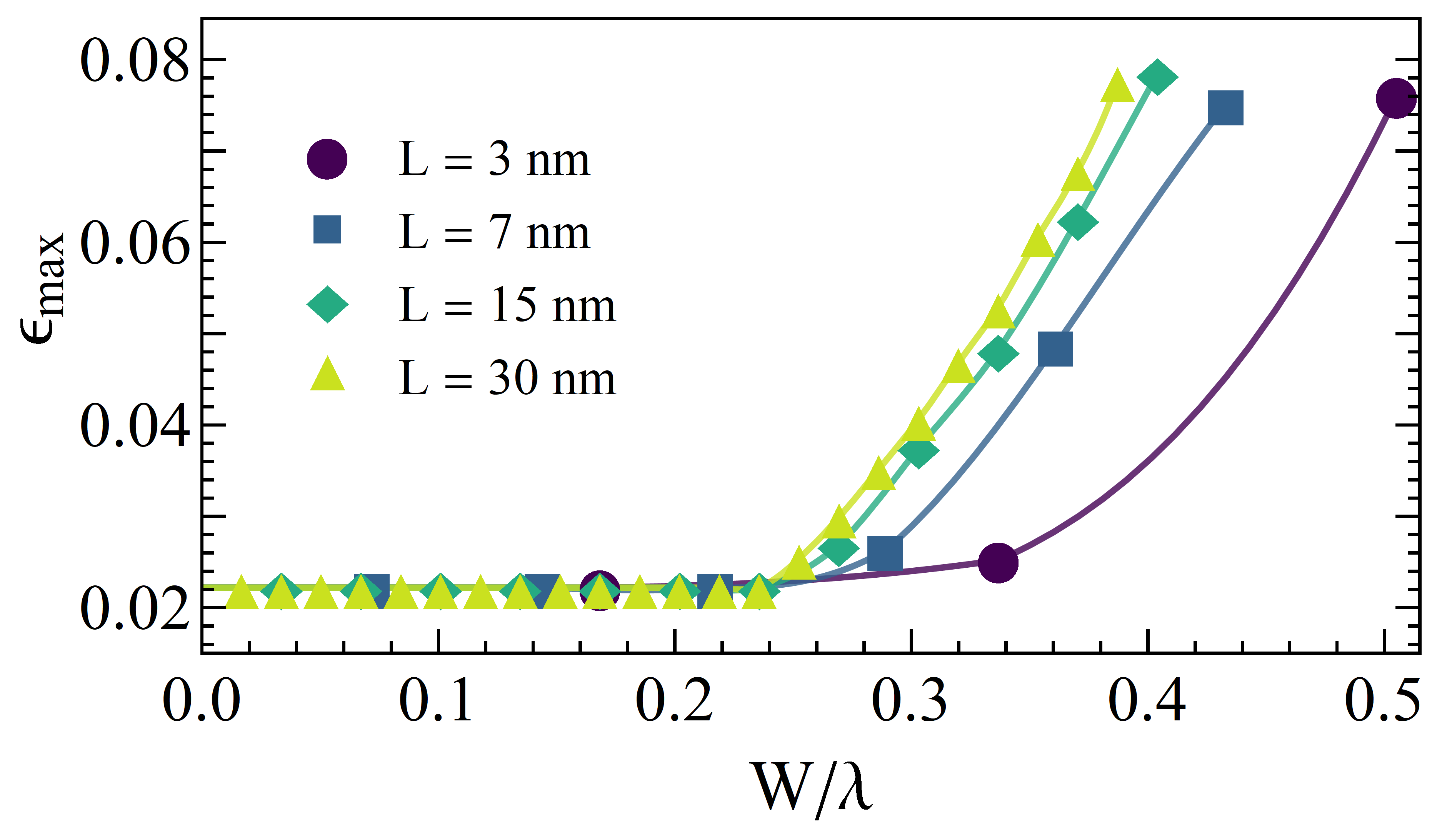}
   \end{center}
   \caption{Maximum value of strain as a function of the torsion coefficient $W/\lambda$ for $3$--ZGNRs having different lengths indicated in the legend. Solid lines are provided as a guide to the eye.}
   \label{fig3}
\end{figure}

Figure~\ref{fig3} shows the maximum strain, $\Epsilon_{\mathrm{max}}$, as a function of the torsion coefficient for $3$--ZGNRs of various lengths. Two regions of different qualitative dependence of the maximum strain on the length can be distinguished. At higher torsion, for $\Upsilon \gtrsim 0.25$, shorter GNRs are less strained than longer ones. These differences are probably related to finite size effects, which is consistent with the fact that the curves tend to a limiting one as the length increases. Within this region, the maximum strain builds up at the edges and grows monotonously with the torsion until it reaches a critical value ($\Epsilon_\mathrm{max} \approx 0.08$) at which some edge bonds break and the configuration of the ribbon becomes irregular. Contrary to that, in the regime of low torsion ($\Upsilon \lesssim 0.25$), the edge bonds are deformed only slightly and the maximum strain builds up at the middle part of the twisted GNR (see the top panel of Figure~\ref{fig2}). More importantly, the maximum strain remains approximately constant (being on the order of $0.02$) and independent on the ribbon length.

So far we have been discussing structural relaxation effects in twisted ZGNRs only. Our simulations showed that twisted AGNRs display more irregular deformation patterns and can generally sustain higher strain. However, as we argue in the next section, AGNRs are less promising from the application point of view and therefore we do not present details of the corresponding relaxation studies.

\section{Electron transmission probability}

%\A{
Modeling of the structural relaxation discussed in the previous section provides complete information on the twisted GNR geometry. In this section we use the computed geometry to study the effects of relaxation on the electron transmission properties of ZGNRs and compare transmission probabilities obtained with and without taking into account the structural relaxation. To this end, on the one hand, we use the computed C atom positions in a relaxed structure to calculate orbital energies $\varepsilon_i$ and \emph{varying} hopping energies $t_{ij}$, defined by Equations (\ref{site-energy-real}) and (\ref{hopping}), respectively, and construct a more realistic Hamiltonian. On the other hand, we build the approximate Hamiltonian using the uniform hopping energy $t_0$ (corresponding to unstrained bonds) and the approximate orbital energies (\ref{site-energy}). 

\begin{figure}[ht]
   \begin{center}
       \includegraphics[width=0.8\linewidth]{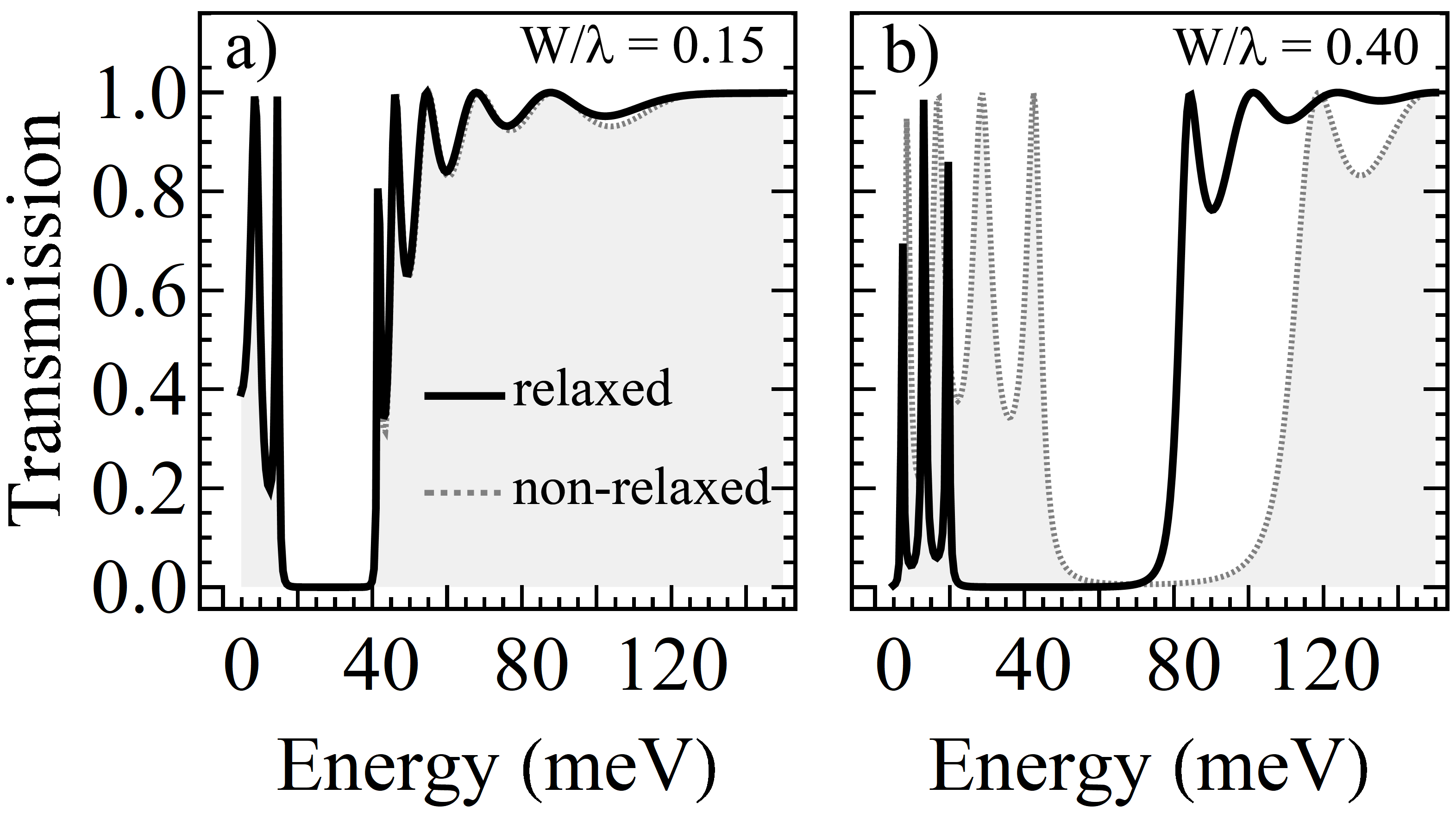}
   \end{center}
   \caption{Transmission coefficient calculated with and without structural relaxation effects taken into account (see text for details) for a $\SI{20}{\nano\meter}$ long $3$--ZGNR, $V_\mathrm{SD}=0$,  $E_z=\SI{25}{\milli\volt/\angstrom}$, and two different values of $W/\lambda$ given in the panels.}
   \label{fig4}
\end{figure}

Then we use the two Hamiltonians to calculate transmission probabilities and compare them. Figure~\ref{fig4} shows a comparison of transmission spectra calculated for a $\SI{20}{\nano\meter}$ long 3--ZGNR, the electric field $E_z=\SI{20}{\milli\volt/\angstrom}$ (fields on this order of magnitude are considered hereafter), and two different values of the torsion coefficient $\Upsilon$. At higher torsion (see the right panel), the transmission spectrum changes substantially when relaxation effects are taken into account. Contrary to that, such changes are negligible for lower torsion coefficient, $\Upsilon=0.15$ (see the left panel). We found that the electron transmission remains almost unaffected in the regime of moderate torsion, $\Upsilon<0.25$. 
%These results are consistent with conclusions of Ref.~[\citenum{Koskinen11}], where it was argued that the electronic structure of a GNR does not change substantially if relaxation effects are taken into account. 
In the rest of the paper, we consider twisted GNRs with $\Upsilon<0.25$ and, therefore, we are using the approximate Hamiltonian for simplicity.

One of the goals of the paper is to propose GNR based nanoelectronic devices in which the electric current can be controlled effectively by minimal operational voltages and fields. In order to find the most sensitive GNR configurations  meeting such requirements, we start by addressing the transmission coefficient at zero bias $V_\mathrm{SD}$ and non-zero transverse electric field $E_z$.  First, we consider a set of GNRs of length $L \simeq \SI{20}{\nano\meter}$ twisted $n=6$ times having different widths and both zig-zag and armchair edges. It is well known that the number of dimer lines $N$ in the transverse direction determines the energy spectrum of AGNRs~\cite{Nakada96,Yang07}. Families of AGNRs with $N=3p$ and $N=3p+1$ ($p$ being a non-negative integer) have a semiconductor-type energy spectra with a wide gap (scaling inversely proportional to the nanoribbon width $W$), while the family with $N=3p+2$ has metallic spectrum. 
%\Ad{reconstruction changes metallic to semiconducting}

\begin{figure}[ht]
   \begin{center}
       \includegraphics[width=0.9\linewidth]{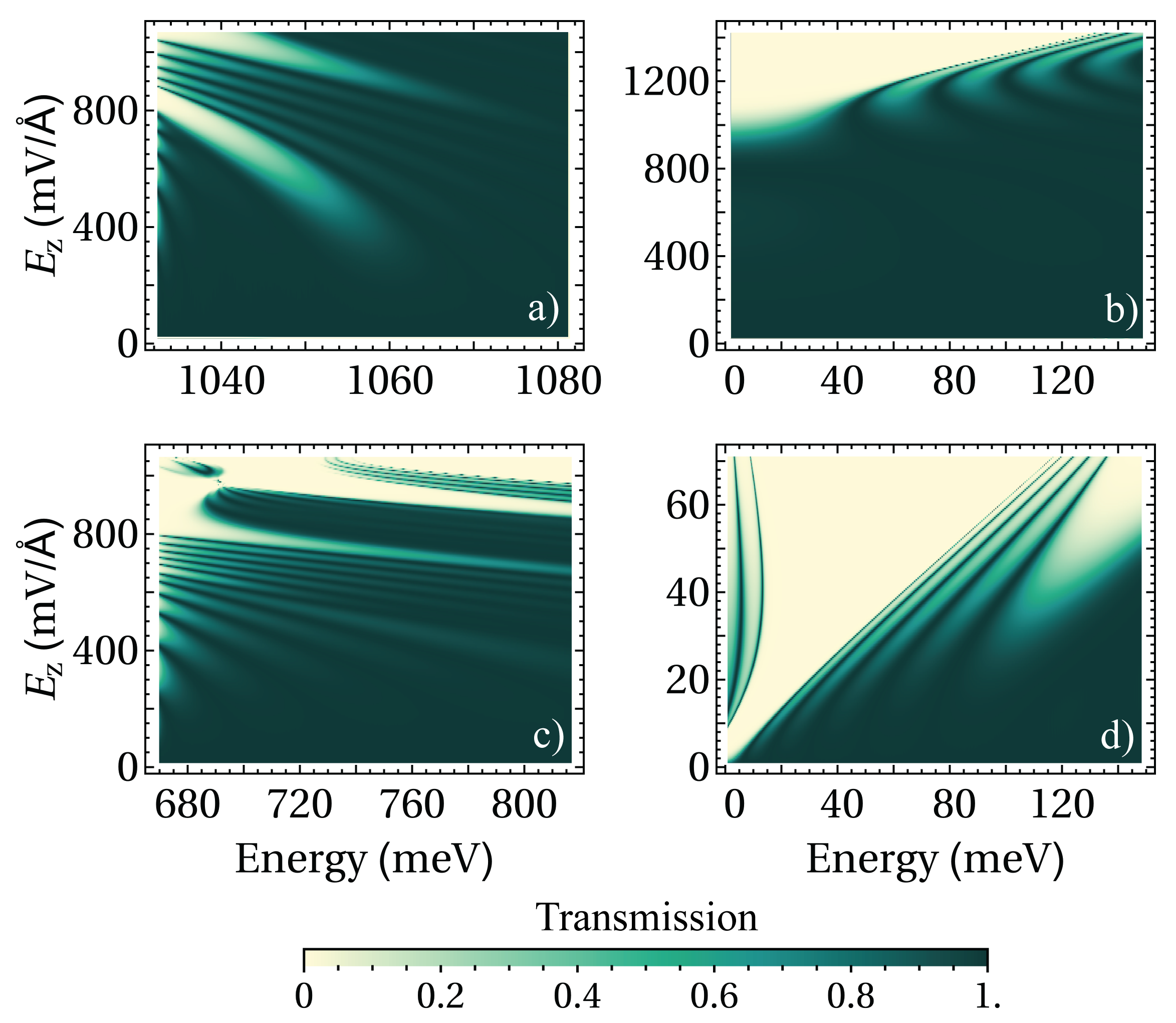}
   \end{center}
   \caption{Maps of the transmission coefficient as a function of the energy and the transverse electric field $E_z$ (at zero bias between the contacts) for several GNRs with $L\simeq \SI{20}{\nano\meter}$ twisted $n=6$ times. (a)~$4$--AGNR (semiconducting), (b)~$5$--AGNR (metallic), (c)~$6$--AGNR (semiconducting), and (d)~$3$--ZGNR (metallic).}
   \label{fig5}
\end{figure}

Figure~\ref{fig5} shows maps of the transmission coefficient as a function of energy and transverse electric field for narrow ribbons of each of the three AGNR families and that for the metallic 3-ZGNR. In each case, the energy range corresponds to the single-mode transmission regime. The figure demonstrates clearly that 
%metallic $5$--AGNR  appears to be the least sensitive to the electric field and it requires higher values of $E_z$ to affect the transmission coefficient noticeably [see Figure~\ref{fig4}(b)]. The latter applies to semiconducting AGNRs too [see Figures~\ref{fig4}(a) and~\ref{fig4}(c)].  On the contrary, 
the control of electron transmission (and consequently the electric current) requires very high values of the transverse field for all AGNRs (see panels (a)-(c) of Figure~\ref{fig5}). Contrary to that, the considered metallic $3$--ZGNR is very sensitive to the controlling transverse electric field, in particular, at low energies [see Figure~\ref{fig5}(d)]. Therefore, we will consider only ZGNRs hereafter. 
%}%\A

\begin{figure}[ht]
   \begin{center}
       \includegraphics[width=0.9\linewidth]{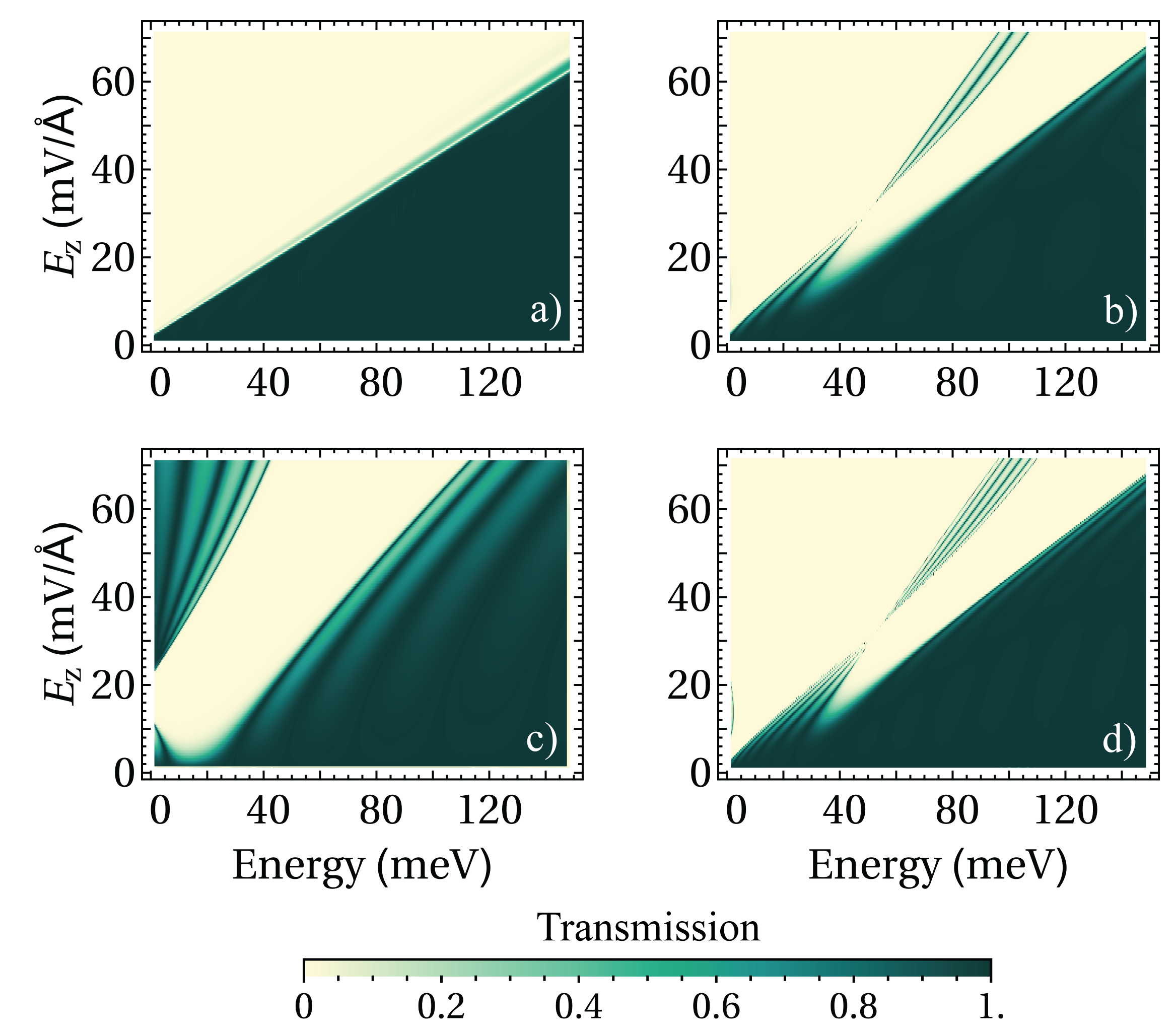}
   \end{center}
   \caption{Maps of the transmission coefficient as a function of the energy and the transverse electric field $E_z$ (at zero bias between the contacts) for several GNRs. The first three panels show results for $3$--ZGNR with $L\simeq \SI{20}{\nano\meter}$ and (a)~$n=1$, (b)~$n=4$, and (c)~$n=10$ respectively. The map for a $3$--ZGNR with $L\simeq \SI{35}{\nano\meter}$ and $n=7$ is given in the panel~(d).} 
   \label{fig6}
\end{figure}

%\A{
Next, we address the influence of the number of twists on the electron transmission of a $3$--ZGNR (with $W\simeq \SI{0.5}{\nano\meter}$). The results are presented in Figure~\ref{fig6}, which shows that even in the case of a single twist [see panel (a)] a gap that is linearly-dependent on the electric field $E_z$ opens in the transmission spectrum. The latter feature can be used for controlling the electric current by the transverse field; such a device would operate as a field-effect transistor. For larger number of twists $n$ [see panels (b) and (d)], additional well isolated lines of high transmission arise in the map; these transmission resonances can be very useful for engineering devices with non-monotonous current-voltage characteristic, as we demonstrate in the next section. Finally, if $n$ is increased even further, the transmission pattern undergoes yet another qualitative change: a gap of zero transmission (a stop band) appears in the spectrum [see Figure~\ref{fig6}(c)]. The parameter controlling the qualitative shape of the transmission pattern is actually not the number of twists but rather the torsion coefficient. To demonstrate this, we compare the transmission spectra of $3$--ZGNRs having different lengths and number of twists but the same value of the torsion coefficient $\Upsilon\simeq 0.1$ [see panels (b) and (d) of Figure~\ref{fig6}]. Despite some expected quantitative differences between the two cases, such as the larger number of resonant lines at larger $n$, the two transmission patterns are qualitatively the same.
%}%\A

\section{Current-voltage characteristics}

% \subsection{Field-effect transistor}
%\A{
Hereafter we study the current-voltage characteristics of $\SI{20}{\nano\meter}$ long $3$--ZGNR (with $W\simeq \SI{0.5}{\nano\meter}$) twisted $n=4$ times. As we have argued above, the dependence of the transmission spectra on the transverse electric field $E_z$ manifests very promising features in this case [see Figure~\ref{fig6}(b)]. Up to now we have been restricting ourselves to the case of zero source drain bias $V_\mathrm{SD}$.  However, for the current-voltage characteristics calculations, it is essential to compute the transmission coefficient taking into account its dependence on the bias $V_\mathrm{SD}$ explicitly, which we do in what follows and then use the Landauer-B\"{u}ttiker formalism to calculate the electric current as~\cite{Datta97}
\begin{equation}
I=\frac{2\,e}{h}\int {\cal{T}}(E,E_z,V_\mathrm{SD}) \Big[f_\mathrm{L}(E)-f_\mathrm{R}(E
%+eV_\mathrm{SD}
)\Big]\,\mathrm{d}E \ ,
\label{I}
\end{equation}
where the Fermi functions of the left and right contacts are given by $f_\mathrm{L}(E) = \left\{1+\exp\left[(\mu-E)/kT)\right]\right\}^{-1}$ and $f_\mathrm{R}(E)=\left\{1+\exp\left[(\mu-e\,V_\mathrm{SD}-E)/kT)\right]\right\}^{-1}$ respectively. Here $\mu$ is the chemical potential at equilibrium, $k$ is the Boltzmann constant, $V_\mathrm{SD}$ is the source-drain voltage (bias) applied across the whole sample in the $x$-direction, and ${\cal{T}}(E,E_z,V_\mathrm{SD})$ is the transmission coefficient depending on energy, transverse field and source-drain voltage. We assume that the chemical potential of both contacts is set to an appropriate point by a back-gate voltage and then the source-drain voltage is applied. All calculations are done for the temperature $T=\SI{4}{\kelvin}$. 

\begin{figure}[ht]
   \begin{center}
       \includegraphics[width=0.6\linewidth]{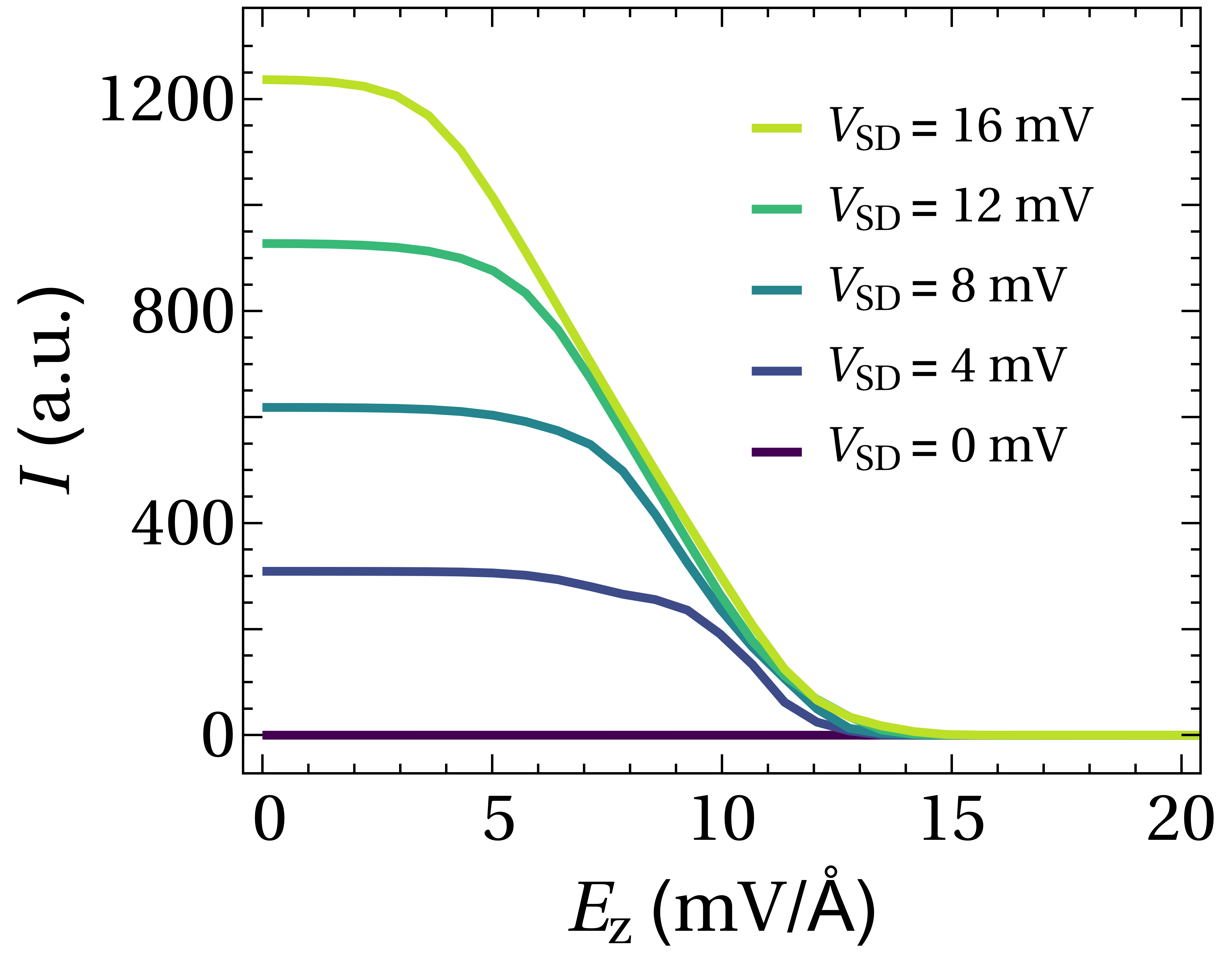}
   \end{center}
   \caption{Electric current $I$ as a function of the controlling transverse electric field, $E_z$, for  $\mu=\SI{20}{\milli\electronvolt}$ and different values of the bias $V_\mathrm{SD}$. The ZGNR geometry as in Figure~\ref{fig6}(b).}
   \label{fig7}
\end{figure}

The dependence of the current on the controlling transverse electric field, $E_z$, calculated for several fixed values of $V_\mathrm{SD}$ is presented in Figure~\ref{fig7}. The figure shows that the electric current can be effectively controlled by the external electric field: the on/off ratio of such a field effect transistor is as high as about $1000$. Provided that the operational point is set appropriately, similar behavior was observed for all ZGNRs we considered, regardless of the dimensions and the number of twists, in particular, in the simplest case of a single twist and $\mu=0$ (not shown here).
%}%\A

\begin{figure}[ht]
   \begin{center}
       \includegraphics[width=0.7\linewidth]{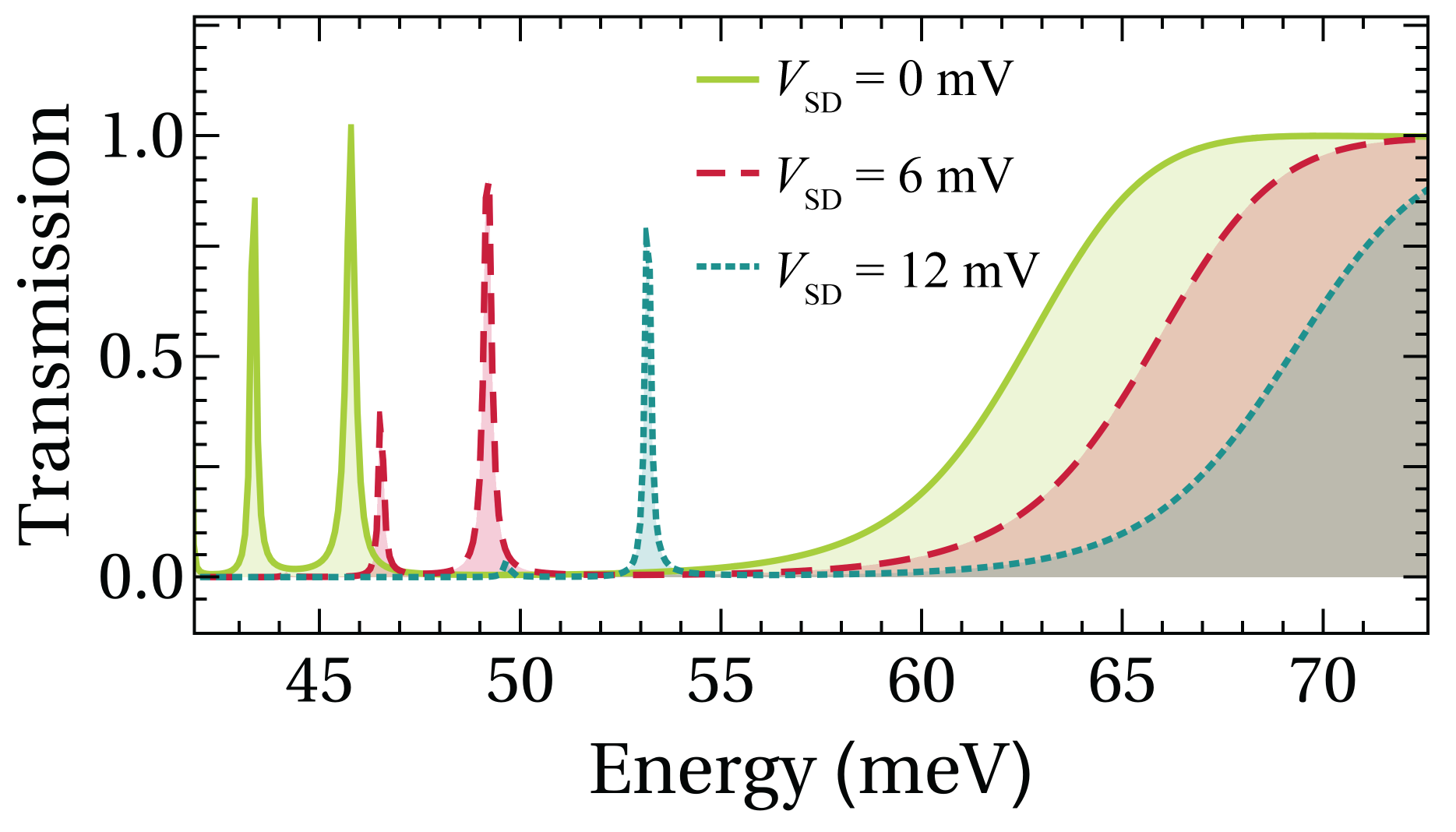}
   \end{center}
   \caption{Transmission coefficient calculated for the fixed value of the transverse electric field, $E_z = \SI{25}{\milli\volt/\angstrom}$, and three different values of $V_\mathrm{SD}$. Resonance peaks shift to higher energy and attenuate as the source-drain voltage increases. The ZGNR geometry is as in Figure~\ref{fig6}(b).}
   \label{fig8}
\end{figure}

\begin{figure}[ht]
   \begin{center}
       \includegraphics[width=0.75\linewidth]{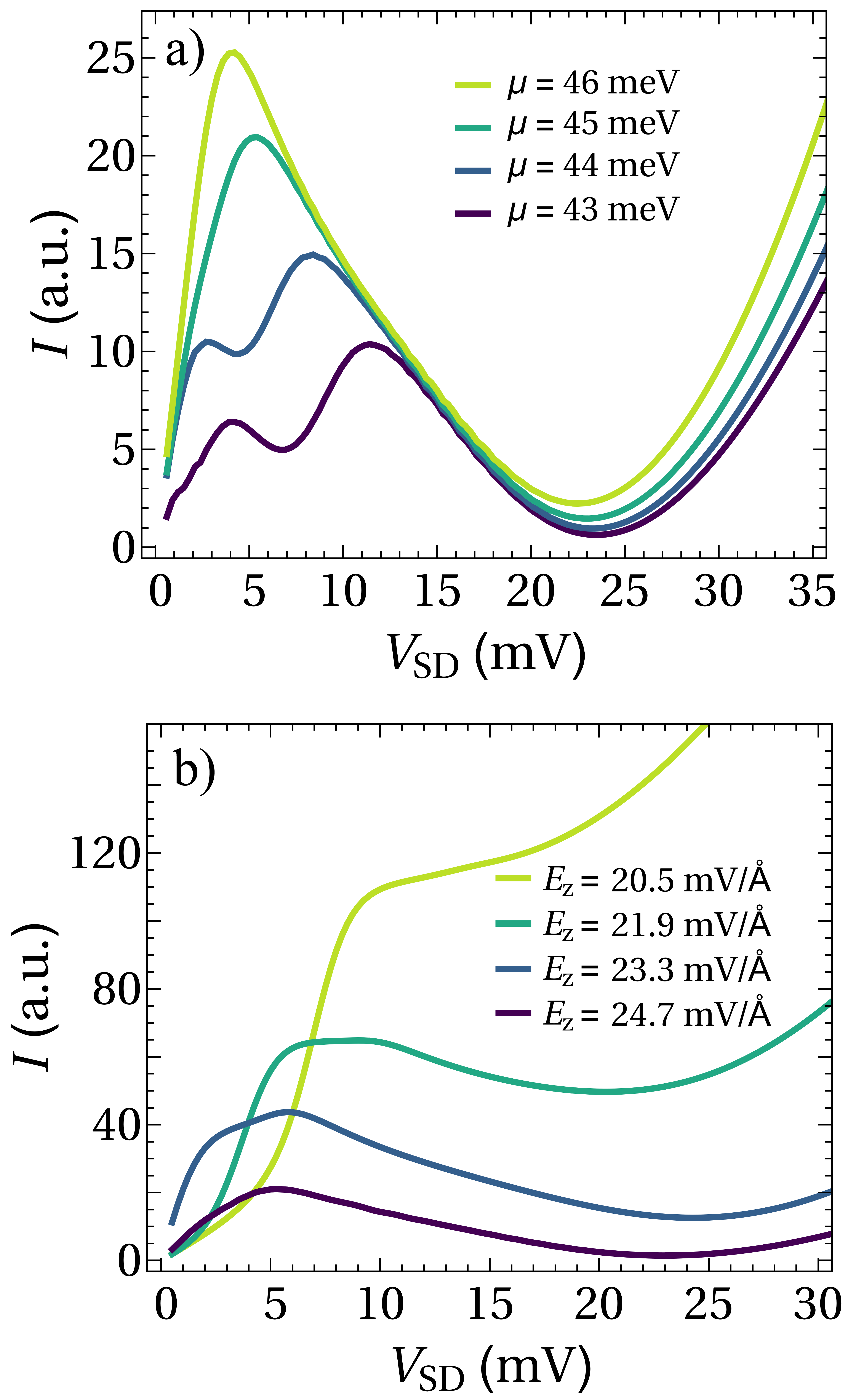}
   \end{center}
   \caption{(a) Current-voltage characteristics, $I(V_\mathrm{SD})$, for different values of the chemical potential, $\mu$, specified in the plot. The electric field is $E_z = \SI{25}{\milli\volt/\angstrom}$. (b) Current-voltage characteristics, $I(V_\mathrm{SD})$, for different values of the applied electric field, $E_z$, specified in the plot. From bottom to top, peak-to-valley ratios are $14.6$, $3.5$ and $1.3$ (the system does not display NDR at the lower value of the electric field $E_z=\SI{20.5}{\milli\volt/\angstrom}$). The chemical potential is $\mu=\SI{45}{\milli\electronvolt}$. For both panels the ZGNR geometry is as in Figure~\ref{fig6}(b).}
   \label{fig9}
\end{figure}

%\subsection{Negative differential resistance}
%\A{
In the most general case the operational point of our GNR based device is determined by the values of the chemical potential $\mu$ and the transverse electric field $E_z$. Below we show that if these parameters are chosen appropriately, the current-voltage characteristics can become N--shaped and have one or several NDR regions. This feature can appear if the transmission spectrum has well defined resonance peaks in the vicinity of the operational point at $V_\mathrm{SD}=0$; see, for example, the straight dark-color inclined lines on the light background in Figure~\ref{fig6}(b). Such resonances can shift and diminish as the source-drain voltage increases, as demonstrated in Figure~\ref{fig8}. Sharp peaks of transmission at zero bias (solid line) shift to higher energies and tend to disappear at larger bias (dashed and dotted curves). The latter decrease in transmission through these resonances at higher bias can give rise to a decrease in the total electric current, resulting eventually in NDR. Indeed, the corresponding current-voltage characteristics have the expected N-shape parts [see Figure~\ref{fig9}(a)]. 

Figure~\ref{fig9}(a) demonstrates that for a fixed magnitude of the transverse field $E_z$, the $I$--$V$ curves have N--shaped parts within a range of values of the chemical potential $\mu$. On the other hand, if the value of $\mu$ is fixed, there is a range of values of the external field $E_z$ for which a NDR region exists in the current-voltage characteristics [see Figure~\ref{fig9}(b)]. The simplest N--shaped $I$--$V$ curves are analogous to those of Gunn diodes~\cite{Gunn1963,Ridley1961} or Esaki tunnel diodes~\cite{Esaki1958,Esaki1960}, suggesting possible applications of twisted ZGNRs as active elements of amplifiers and generators. The traditional figure of merit of the latter devices is the peak-to-valley current ratio in the NDR region, which can be controlled in the case of twisted GNRs by adjusting the operational point. Moreover, as the figure shows, one can engineer also $I$--$V$ curves with at least two NDR regions by varying the controlling parameters. The latter is opening a possibility of new classes of digital applications: it has a potential to go beyond conventional binary logic by using several overlapping NDR regions to obtain multiple stable logic states. Thus, the underlying characteristics of GNR based nanoscopic devices are tunable by external macroscopic parameters. 
%}%\A

\section{Conclusions} \label{sec:conclusions}

In conclusion, we have studied the electronic transport properties of twisted gra\-phene nanoribbons subjected to an external transverse electric field. By means of the density-functional based tight-binding method, we showed that effects of the twist-induced strain on the transmission spectrum are negligible within a wide range of values of the torsion deformation. We demonstrated that our proposed simplified tight-binding model with constant hopping energy gives reliable results in relevant cases, suggesting that our model can be used instead of more computationally intensive methods. We argued that twisted GNRs with zig-zag edges are more promising for applications since their transmission characteristics are highly sensitive to the transverse electric field even at low values of the field. Thus, the source-drain current in a twisted ZGNR can be effectively controlled by the external field; in this case the system operates as a field-effect transistor with the on/off ratio on the order of $1000$. We demonstrate also that if the operational point is set appropriately, twisted ZGNRs have current-voltage characteristics which are tunable by the transverse electric field; in this way $I$--$V$ curves can be engineered to have one or several NDR regions with multiple stable states. Our findings suggest a number of potential applications in graphene-based nanoelectronics, such as field-effect transistors, active elements of amplifiers and generators, and new generation of logic elements with multiple logic states, which go beyond conventional binary logic.

%\F{???}
%Use of multiple NDR regions used to achieve multiple logic states may result in highly dense devices and fewer circuit elements than prior generations

%\begin{acknowledgements}
%\section{Acknowledgements}
\acknowledgments

This work has been supported by MINECO (Grant MAT2016-75955). M. S.-B. and F. D.-A. thank L.~Medrano and R.~Gutierrez for helpful conversations.

%\end{acknowledgements}

\bibliography{references}

\end{document}